\documentclass[%
 preprint, linenumbers,
 amsmath,amssymb,
 aps, physrev,
]{revtex4-2}

\usepackage{graphicx}
\usepackage{dcolumn}
\usepackage{bm}

\usepackage{newtxtext}
\usepackage{newtxmath}
\usepackage{natbib}

\usepackage{epstopdf, epsfig}
\usepackage{comment}
\usepackage{tabularx}
\usepackage{lipsum}
\usepackage{amssymb}
\usepackage{amsmath}
\usepackage{overpic}
\usepackage{epstopdf}
\usepackage[T1]{fontenc}
\usepackage[utf8]{inputenc}
\usepackage{color,soul}
\usepackage{placeins}
\usepackage{hyperref}
\usepackage{caption}

\usepackage[utf8]{inputenc}

\DeclareMathSymbol{\shortminus}{\mathbin}{AMSa}{"39}

\hypersetup{
    colorlinks = true,
    urlcolor   = blue,
    citecolor  = black,
}

\newcommand{\RomanNumeralCaps}[1]

\begin{document}

\nolinenumbers

\preprint{APS/EKI2D}

\title{\textbf{Online learning of eddy-viscosity and backscattering closures for geophysical turbulence using Ensemble Kalman Inversion} 
}%

\author{Yifei Guan}
 \email{Contact author: guany@union.edu}
\affiliation{Department of Mechanical Engineering, Union College, Schenectady, NY, USA.\\
Department of Geophysical Sciences, The University of Chicago, Chicago, IL, USA}

 
\author{Pedram Hassanzadeh}%
 \email{Contact author: pedramh@uchicago.edu}
\affiliation{%
Department of Geophysical Sciences and Committee on Computational and Applied Mathematics, The University of Chicago, Chicago, IL, USA
}%

\author{Tapio Schneider}
\author{Oliver Dunbar}%
\affiliation{%
 Division of Geological and Planetary Sciences, California Institute of Technology, Pasadena, CA, USA
}%

\author{Daniel Zhengyu Huang}
\affiliation{
 Beijing International Center for Mathematical Research, Peking University, Beijing, China
}%

\author{Jinlong Wu}
\affiliation{%
 Department of Mechanical Engineering, University of Wisconsin–Madison, Madison, WI, USA
}%

\author{Ignacio Lopez-Gomez}
\affiliation{%
 Google Research,  Mountain View, CA, USA
}%


\date{\today}

\begin{abstract}
Different approaches to using data-driven methods for subgrid-scale closure modeling of geophysical turbulence have emerged recently. Most of these approaches are data-hungry, and lack interpretability and out-of-distribution generalizability. Here, we use a hybrid approach that combines turbulence theory, physics-based modeling, and data-driven methods to overcome these challenges. Specifically, we address the parametric uncertainty of well-known physics-based large-eddy simulation (LES) closures: the Smagorinsky (Smag) and Leith eddy-viscosity models (1 free parameter) and the Jansen-Held (JH) backscattering model (2 free parameters). For 8 cases of 2D geophysical turbulence, optimal parameters are first learned online from data via ensemble Kalman inversion (EKI), such that for each case, the LES' energy spectrum matches that of direct numerical simulation (DNS). Only a small training dataset is needed (to calculate the DNS spectra); i.e., the approach is {data-efficient}. We find the optimized parameter(s) of each closure to be constant across broad flow regimes that differ in dominant length scales, eddy/jet structures, and dynamics, suggesting that these closures are {generalizable}. Next, we show that the online learned constants agree with the predictions of a recent semi-analytical derivation, providing further interpretability. In both a-priori and a-posteriori tests, that include examining the extreme events, LES with optimized closures, especially with JH, outperform the baselines (LES with standard Smag, dynamic Smag or Leith). This work shows the promise of combining advances in theory, physics-based modeling (e.g., JH), and data-driven modeling (e.g., {online} learning with EKI) to develop data-efficient frameworks for accurate, interpretable, and generalizable closures for geophysical turbulence, with ultimate applications in weather and climate prediction. 
\end{abstract}

\maketitle

\section{Introduction}\label{sec:Introduction}
Recent advances in data-driven methods offer new avenues to develop better subgrid-scale (SGS) closures (parameterizations), which are essential, for example, for large-eddy simulations (LES) of geophysical turbulence in critical applications such as climate modeling \cite[e.g.,][]{mansfield2022calibration,schneider2023harnessing,sanderse2024scientific,lai2024machine,eyring2024pushing,bracco2025machine,mansfield2024uncertainty}. Most studies have focused on \textit{offline} (supervised) learning, in which a closure is developed, for instance, by training neural networks (NNs) or using equation-discovery algorithms to match the true and parameterized SGS fluxes \cite[e.g.,][]{maulik2019subgrid,beck2019deep, zanna2020data, guan2021stable, jakhar2023learning}. The trained/discovered closure, once validated {\it a-priori}, is coupled to a low-resolution solver (e.g., a climate model) for {\it a-posteriori} tests. While examples of stable and accurate SGS modeling using \textit{offline} learning with NNs have emerged recently \cite[e.g.,][]{guan2023learning,srinivasan2023turbulence,ross2023benchmarking,yuval2023neural}, this approach has major drawbacks: 1) training NNs require many samples of the ``true'' SGS fluxes, 2) these closures are largely uninterpretable, and 3) they do not generalize out-of-distribution (extrapolate) to other flows (e.g., higher Reynolds numbers, $Re$); see, e.g., \cite{rasp2018deep, subel2023explaining, beucler2024climate}. Item (1) is a major practical challenge as it requires long, expensive, high-resolution simulations (e.g., direct numerical simulations, DNS). Also, estimating the true SGS fluxes from such data is difficult \cite[]{jakhar2023learning,sun2023quantifying}. 

An alternative is \textit{online} (aka end-to-end) learning: the closure is developed while it is coupled to the low-resolution solver, with the objective of matching certain statistics of the ``true'' flow (e.g., from high-resolution simulations or even observations). If the existing closures have ``structural uncertainties'', then NNs can be used to learn either the entire closure from data~\cite[e.g.][]{Kennedy2002Bayesian} or corrections to physics-based closures~\cite[e.g.][]{wu2024learning}. However, minimizing the loss function of this approach can be practically challenging (see below); that said, promising results with NNs trained using ensemble Kalman inversion (EKI) \cite[]{pahlavan2024explainable}, multi-agent reinforcement learning \cite[]{novati2021automating,mojgani2023extreme}, ensemble-based synchronization~\cite[]{buzzicotti2020synchronizing,ruiz2013estimating}, and differentiable modeling \cite[]{sirignano2020dpm,frezat2022posteriori,list2022learned, shankar2023differentiable} have started to appear. While this approach addresses item (1), interpretability and generalization ((2)-(3)) remain major challenges.

One avenue for addressing (1)-(3) is to develop closures with the appropriate structure using physical arguments (and when needed, equation-discovery techniques), and use \textit{online} learning to estimate the unknown parameters (i.e., addressing ``parametric uncertainty''); see \cite{schneider2021accelerating} for further discussions and \cite{schneider2023harnessing,schneider2024opinion},  \cite{lopez2022training}, \cite{zhang2022ensemble}, and \cite{martinez2024relaxation} for recent examples of success with this approach. The major question about this approach is whether one is simply calibrating/tuning a closure for limited metrics, such that the calibrated closure does not work well for other metrics. Furthermore, generalization (3) requires the estimated parameters to be universal constants (or at least nearly invariant within the flow regimes of interest). These questions, even for canonical test cases, remain to be fully investigated.  



In this paper, we aim to answer these questions for several setups of 2D turbulence, a canonical model for many geophysical flows~\cite[e.g.,][]{boffetta2012two,davidson2015turbulence,vallis2017atmospheric, gallet2020vortex}. We estimate the parameters of two classical eddy-viscosity closures (Smagorinsky and Leith) and a new backscattering closure \cite[]{jansen2014parameterizing} \textit{online} using EKI and by matching the energy spectrum of LES with that of a short DNS (truth).
The LES with EKI-optimized parameters, which are found to agree with a recent semi-analytical model for these parameters, are then examined in terms of enstrophy spectra, interscale transfers, and probability density functions (PDFs) of vorticity, especially at the tails (extreme events). 

\section{Data and Methods}\label{sec:Method}   


\subsection{DNS and LES of 2D turbulence}
\begin{figure}
\vspace{.0in}
 \centering
 \vspace*{2mm}
 \begin{overpic}[width=1\linewidth,height=0.5\linewidth]{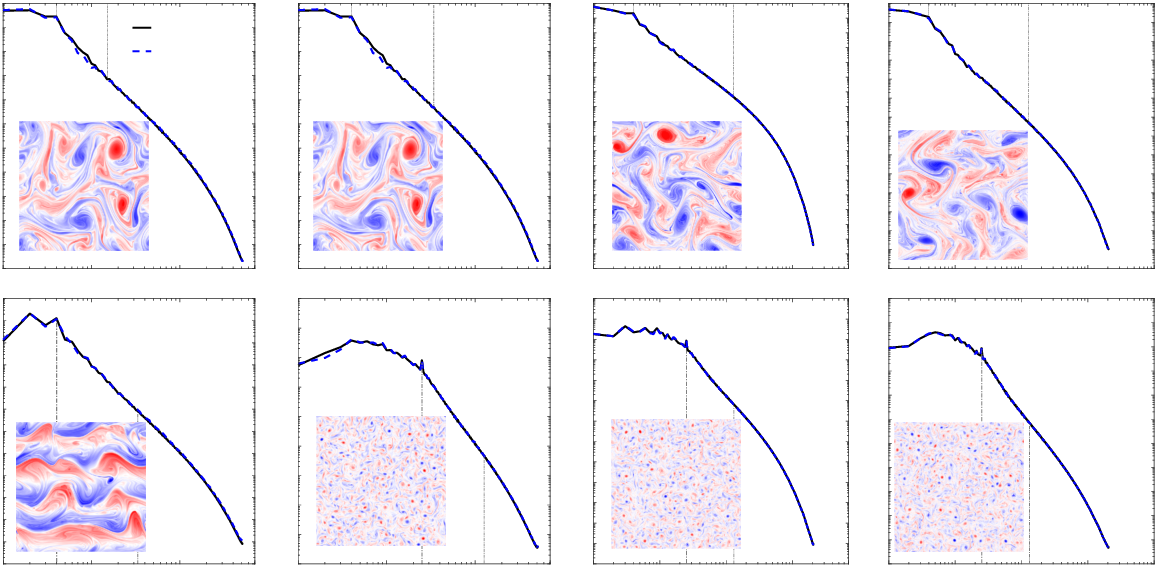}

 \put(13.2,47){\scriptsize{DNS}}
 \put(13.2,45){\scriptsize{DNS$^*$}}

 \put(-5,-2){\scriptsize{$k=$}}

 \put(-1,-2){\scriptsize{$10^0$}}
 \put(6.5,-2){\scriptsize{$10^1$}}
 \put(14,-2){\scriptsize{$10^2$}}

 \put(24.5,-2){\scriptsize{$10^0$}}
 \put(32,-2){\scriptsize{$10^1$}}
 \put(40,-2){\scriptsize{$10^2$}}

 \put(50,-2){\scriptsize{$10^0$}}
 \put(56,-2){\scriptsize{$10^1$}}
 \put(61.5,-2){\scriptsize{$10^2$}}
 \put(67.5,-2){\scriptsize{$10^3$}}

 \put(75.5,-2){\scriptsize{$10^0$}}
 \put(81,-2){\scriptsize{$10^1$}}
 \put(87,-2){\scriptsize{$10^2$}}
 \put(93,-2){\scriptsize{$10^3$}}

 \put(4,50.8){\scriptsize{$k_f$}}
 \put(8.2,50.8){\scriptsize{$k_{c}$}}

 \put(29.5,50.8){\scriptsize{$k_f$}}
 \put(37,50.8){\scriptsize{$k_{c}$}}

 \put(53.8,50.8){\scriptsize{$k_f$}}
 \put(63,50.8){\scriptsize{$k_{c}$}}

 \put(79.5,50.8){\scriptsize{$k_f$}}
 \put(88,50.8){\scriptsize{$k_{c}$}}

 \put(2.5,18){\scriptsize{$k_f$}}
 \put(12,14){\scriptsize{$k_{c}$}}

 \put(36,19){\scriptsize{$k_f$}}
 \put(41.5,11){\scriptsize{$k_{c}$}}

 \put(59,20.5){\scriptsize{$k_f$}}
 \put(63.5,15){\scriptsize{$k_{c}$}}
 
 \put(84,21){\scriptsize{$k_f$}}
 \put(89,15){\scriptsize{$k_{c}$}}

  \put(2, 40.5){\scriptsize{Case 1.1}}
  \put(38.5, 45){\scriptsize{Case 1.2}}
  \put(64.5, 45){\scriptsize{Case 1.3}}
  \put(91, 45){\scriptsize{Case 1.4}}

  \put(15.0, 20){\scriptsize{Case 2}}
  \put(38.5, 20){\scriptsize{Case 3.1}}
  \put(64.5, 20){\scriptsize{Case 3.2}}
  \put(91.0, 20){\scriptsize{Case 3.3}}

  \put(-1, 49.5){\tiny $0$}
  \put(-2.5, 26){\tiny $\shortminus 11$}
  \put(24.5, 49.5){\tiny $0$}
  \put(23.0, 26){\tiny $\shortminus 11$}
  \put(50, 49){\tiny $0$}
  \put(48.5, 26){\tiny $\shortminus 18$}
  \put(75.0, 49){\tiny $0$}
  \put(74.0, 26){\tiny $\shortminus 14$}

  \put(-1, 23){\tiny $1$}
  \put(-2.5, 0){\tiny $\shortminus 11$}
  \put(24.5, 23){\tiny $0$}
  \put(23.8, 0){\tiny $\shortminus 9$}
  \put(50, 23){\tiny $0$}
  \put(48.3, 0){\tiny $\shortminus 13$}
  \put(75.5, 23){\tiny $0$}
  \put(74.0, 0){\tiny $\shortminus 11$}

 \end{overpic}
  \vspace*{0mm}
  \caption{\footnotesize The TKE spectra ($\text{log}_{10}(\hat{E}(k))$) and examples of $\omega$ for each case. The forcing wavenumber $k_f$ and cut-off wavenumber $k_c$ are marked. The black (blue) lines show DNS spectra averaged over all 100 (first 5) snapshots in the training set. With a sharp-spectral cut-off filter, the FDNS spectrum, which is used as the target in Eq.~\eqref{eq:loss}, overlaps with the DNS spectrum over $k=[0,k_c]$. }
 \label{fig:2 DNS and FDNS}
 \vspace*{0mm}
\end{figure}
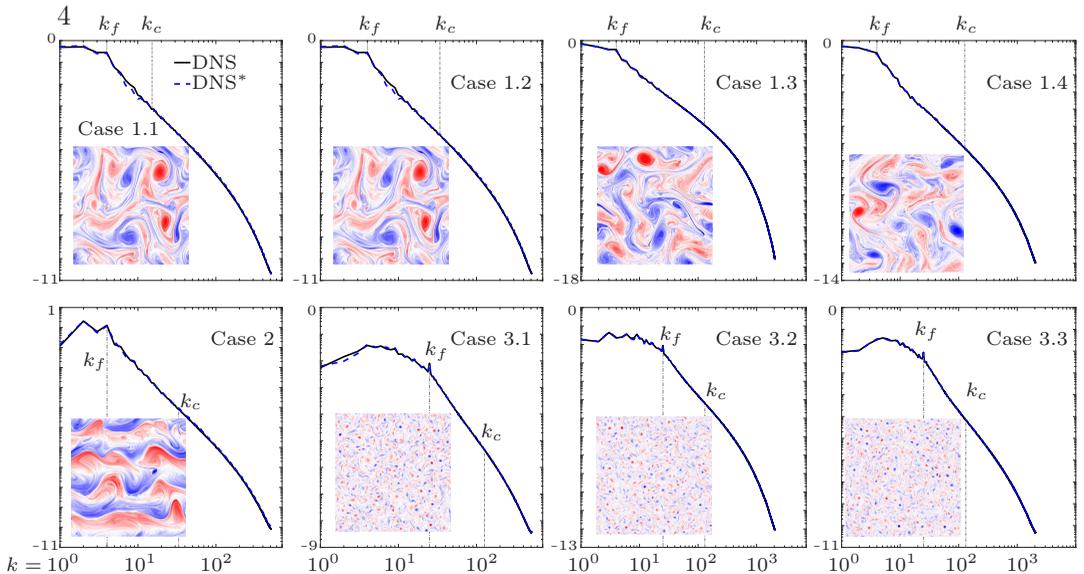

We use 8 setups of forced 2D $\beta$-plane turbulence. The dimensionless governing equations in the vorticity ($\omega$) and streamfunction ($\psi$) formulation in a doubly periodic square domain of length $L=2\pi$ are
\begin{eqnarray}\label{eq:NS}
\frac{\partial \omega}{\partial t} + \mathcal{N}(\omega,\psi)=\frac{1}{Re}\nabla^2\omega - f -r\omega + \beta\frac{\partial \psi}{\partial x}, \;\nabla^2\psi = -\omega.
\end{eqnarray}
Here, $\mathcal{N}$ is the Jacobian, $f(x,y)= k_f[\cos{(k_fx)} + \cos{(k_fy)}]$ is a deterministic forcing, and $r=0.1$ is the linear drag coefficient (e.g., representing surface friction); $\beta$ is the gradient of the Coriolis parameter. We study 8 cases, in which $Re$, forcing wavenumber ($k_f$), and $\beta$ have been varied (Table~\ref{tab:1}), creating a variety of flows that differ in dominant length scales and energy/enstrophy cascade regimes (see Fig.~\ref{fig:2 DNS and FDNS}). For DNS, which is treated as the truth, Eqs.~\eqref{eq:NS} are solved at high spatial-temporal resolutions ($N^2_\text{DNS} \times \Delta t_\text{DNS} $) using a Fourier pseudo-spectral solver~\cite[see][]{guan2021stable}. The LES equations are derived by applying a low-pass filter (e.g., sharp cut-off), denoted by $\overline{(\cdot)}$, to \eqref{eq:NS}:
\begin{eqnarray}\label{eq:FNS}
\frac{\partial \overline{\omega}}{\partial t} + \mathcal{N}(\overline{\omega},\overline{\psi})=\frac{1}{Re}\nabla^2\overline{\omega}-\overline{f}-r\overline{\omega}+\beta\frac{\partial \overline{\psi}}{\partial x}-\underbrace{\left[\overline{\mathcal{N}({\omega},{\psi})}-\mathcal{N}(\overline{\omega},\overline{\psi})\right]}_{\Pi^\text{SGS}=\nabla\times(\nabla \cdot \tau^\text{SGS})}, \; \nabla^2\overline{\psi} = -\overline{\omega}.
\end{eqnarray}
Solving these equations needs much coarser resolutions ($N^2_\text{LES} \times 10\Delta t_\text{DNS}$), where $N_\text{DNS}/N_\text{LES}$ is between 4 and 32 (Table~\ref{tab:1}). The SGS term, $\Pi^\text{SGS}$ or ${\tau}^{\text{SGS}}$, requires a closure.

\begin{table}
\renewcommand{\arraystretch}{1.5}
\vspace*{4mm}
{\scriptsize
  \begin{center}

\begin{tabular}{|c|c|c|c|c|c|c|c|c|}
     \hline
     Case & 1.1 & 1.2 & 1.3 & 1.4& 2 & 3.1 & 3.2 & 3.3\\ \hline
       $Re$   &20000 &20000 & 100000 & 300000 & 20000 & 20000 & 100000 & 300000\\
       $k_f$  &4 &4&4 &4 &4 &25&25 &25 \\
       $\beta$ &0&0&0&0&20&0&0&0 \\
       $N_\text{DNS}$ &1024&1024&4096&4096&1024&1024&4096&4096 \\
       $N_\text{LES}$ &32&64&256&256&64&256&256&256  \\ \hline \color{black}$A$&1.87&1.87&1.85&1.81&2.48&1.88&1.77&1.79 \\
       \hline
       $C_{\text{S}}^{\text{EKI}}$   & {0.12}& {0.12}& {0.11}& {0.12}& 0.10& {0.12}& {0.12}& {0.10}\\

       \color{black} $C_{\text{S}}^{\text{Analytical}}$   &0.13&0.12&0.11&0.12&0.10&0.11&0.12&0.12\\
       
       $C_{\text{L}}^{\text{EKI}}$   &{0.23}&{0.25}&{0.26}&{0.24}&0.21&{0.24}&{0.23}&{0.21}\\
       
       \color{black} $C_{\text{L}}^{\text{Analytical}}$   &0.23&0.23&0.23&0.24&0.20&0.23&0.24&0.24\\
       
       $C_{\text{JHS}}^{\text{EKI}}$, $C_{\text{B}}^{\text{EKI}}$   &{0.23, 0.96}&{0.22, 0.95}&{0.21, 0.95} & {0.21, 0.95}&0.20, {0.94}&{0.22, 0.94}&0.23, {0.95}&{0.21, 0.96} \\ 
       
       $C_{\text{JHL}}^{\text{EKI}}$, $C_{\text{B}}^{\text{EKI}}$   &0.34, {0.95}&{0.32, 0.95}&{0.33, 0.94}&0.30, {0.94}&{0.31, 0.96}&{0.32, 0.95}&{0.32, 0.94}&{0.31, 0.93} \\ 
       
       \color{black} $C_{\text{JHL}}^{\text{Analytical}}$, $C_{\text{B}}^{\text{Analytical}}$   &0.35, 0.95&0.34, 0.95&0.33, 0.94&0.33, 0.94&0.31, 0.96&0.33, 0.95&0.34, 0.94&0.33, 0.93 \\ 
    \hline
  \end{tabular}
  \vspace*{0mm}
  \captionof{table}{Top: Physical and numerical parameters of the 8 cases. Bottom: EKI-optimized and {\color{black} semi-analytically-derived} parameters (uncertainties are shown in Supplementary Material Table~I). The value of $A$ is diagnosed from the inertial region of the DNS TKE spectrum for each case. The semi-analytical values {\color{black}$C_{\text{S}}^{\text{Analytical}}$, $C_{\text{L}}^{\text{Analytical}}$, and $C_{\text{JHL}}^{\text{Analytical}}$ are given by Eqs.~\eqref{eq:CS},~\eqref{eq:CL} and \eqref{eq:CHL_CB}, and $C_{\text{B}}^{\text{Analytical}}$ is chosen to be the same as $C_{\text{B}}^{\text{EKI}}$ for JHL.}}
  \label{tab:1}
  \end{center}
  \vspace*{1mm}
  }
\end{table}
\subsubsection{Physics-based closures with parametric uncertainty}
Eddy-viscosity closures are ``functional'' models that assume the net effect of the SGS term is dissipation of the resolved scales: ${\tau}^{\text{SGS}} = -2\nu_e \bar{\mathcal{S}}$, where $\bar{\mathcal{S}}$ is the rate of strain of the resolved flow, and $\nu_e(x,y,t)$ is the eddy-viscosity~\citep[]{sagaut2006large}. For 2D turbulence, the Smagorinsky model (Smag: $\nu_e = (C_{\text{S}}\Delta)^2 \langle\bar{\mathcal{S}}^2\rangle^{1/2}$, $\Delta=L/N_\text{LES}$)~\cite{smagorinsky1963general} and the Leith model (Leith: $\nu_e = (C_{\text{L}}\Delta)^3 {\langle(\nabla \bar{\omega})^2\rangle^{1/2}}$)~\cite{leith1996stochastic} are often used in practice, with $C_{\text{S}}$ and $C_{\text{L}}$ being  model parameters to be determined ($\langle \cdot \rangle$ means domain averaging). For 3D homogeneous isotropic turbulence, a constant $C_{\text{S}}=0.17$ was analytically derived from the energy spectrum scaling~\cite[][]{lilly1967representation,pope2001turbulent}. Empirically, it was found that $0.17$ leads to the best {\it a-priori} performance~\cite[]{mcmillan1979direct}. For 2D turbulence, no analytical or empirical estimate existed for $C_{\text{S}}$ or $C_{\text{L}}$, and trial and error was often used to select a value \cite[]{maulik2019subgrid} until very recently, when a semi-analytical estimate for $C_{\text{S}}$ and $C_{\text{L}}$ was presented in~\cite{guan2025semi}. For both closures, $C$ can be also determined dynamically (DSmag, DLeith) from the Germano identity~\cite[][]{germano1991dynamic,khani2015large,grooms2023backscatter}. However, this procedure can lead to $\nu_e<0$ (backscattering) and unstable LES. While capturing backscattering is desirable (see below), in practice, DSmag and DLeith are used with ``positive clipping'', which enforces $\nu_e\geq0$ and makes the closure more diffusive (thus less accurate) in favor of stability.  

To capture backscattering, which can be important in many flows \cite[]{sagaut2006large}, including oceanic and atmospheric circulations~\cite[]{khani2016backscatter,juricke2020kinematic,grooms2023backscatter}, backscattering closures have been developed that, for example, inject a fraction of the dissipated energy back to the resolved scales. Here, we use the recently developed closure of \cite{jansen2014parameterizing} (JH, hereafter), which has been applied to oceanic and quasi-geostrophic turbulence~\cite[][]{jansen2015energy,ross2023benchmarking}. The JH closure is
\begin{eqnarray} \label{eq:JH}
{\Pi}^{\text{SGS}} = \nabla^2(\nu_e \nabla^2\bar{\omega})+\nu_b\nabla^2\bar{\omega}, \; \text{ }\nu_b = -C_{\text{B}}\left\langle \bar{\psi}\nabla^2(\nu_e \nabla^2\bar{\omega})\right\rangle/\langle\bar{\psi}\nabla^2\bar{\omega}\rangle,
\end{eqnarray}
where the first term is biharmonic eddy-viscosity, and the second term represents backscattering with anti-diffusion. $C_\text{B}$ determines how much of the globally dissipated energy is fed back into the resolved scales. $\nu_e$ can be determined in the same way as in Leith \cite[][]{jansen2014parameterizing}) or Smag~\cite[][]{jansen2015energy,ross2023benchmarking}. Like $C_\text{S}$ and $C_\text{L}$, $C_{\text{B}}$ is an empirically determined parameter that is expected to be $\mathcal{O}(1)$ to balance the forward transferred and backscattered energy. JH showed that in general, $C_{\text{B}}\geq 0.9$, and used $0.9$ in their research~\cite{jansen2014parameterizing}. 


Here, we focus on four closures: Smag, Leith, JH-Smag (JHS) and JH-Leith (JHL). We estimate, as described below, their one or two \textit{constant} parameters $C$ for each case. 
For JHS and JHL, we define $\nu_e$ in similar ways as in Smag/Leith \cite[]{jansen2014parameterizing,jansen2015energy,ross2023benchmarking} but choose the power of $C\Delta$ consistent with the dimension of the biharmonic: $\nu_e = (C_{\text{JHS}}\Delta)^4\langle\bar{\mathcal{S}}^2\rangle^{1/2}$ (JHS) and $\nu_e=(C_\text{JHL}\Delta)^6\langle(\nabla^2 \bar{\omega})^2\rangle^{1/2}$ (JHL). Note that to be consistent with other implementations, unlike JH14 or JH15, we use domain averaging in the calculation of $\nu_e^\text{JHL}$. Here, like $C_{\text{S}}$ and $C_{\text{L}}$, parameters $C_{\text{JHS}}$, $C_{\text{JHL}}$, and $C_{\text{B}}$ are dimensionless.

\subsection{Calibration, emulation, and sampling (CES) for {\it online} learning}
\begin{figure}
 \centering
 \begin{overpic}[width=1.0\linewidth,height=0.4\linewidth]{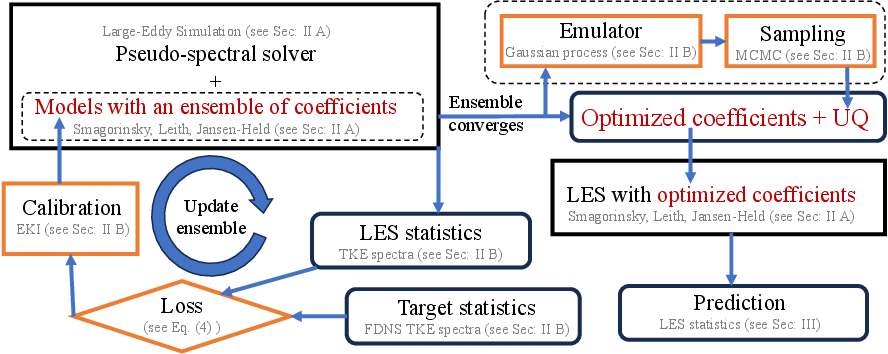}
 \end{overpic}
\vspace*{-4.0mm} \caption{\footnotesize The schematic of the CES framework with EKI to estimate the closure parameters with uncertainty. Briefly, first, for a given case and closure, an ensemble of $C$ values are generated by drawing 10 times from a Gaussian distribution (see Supplementary Material). For each $C$ value(s), a long-enough LES is performed and the mean TKE spectrum of this ensemble member is computed. Then the loss is calculated against the target (FDNS spectrum) for each ensemble member. EKI is used to update the $C$ values, and this process continues until the ensembles converge, which usually happens in 10 or fewer epochs (a typical convergence trend is shown in Supplementary Material Fig.~1). For uncertainty quantification (UQ), all parameters and their corresponding statistics from all epochs are used to train a Gaussian process emulator. This fast emulator is used for UQ with Markov-chain Monte Carlo (MCMC).} 
 \label{fig:1 framework}
 \vspace*{-4.0mm}
\end{figure}

The full calibration, emulation, and sampling (CES) framework applied in this work is depicted in Fig.~\ref{fig:1 framework}. We estimate the optimal $C$ parameter(s) (e.g., $C_\text{S}$, $C_\text{L}$, $C_\text{JHS}$, $C_\text{JHL}$, $C_\text{B}$) of a closure by matching the TKE spectra $(\hat{E}(k))$ of the filtered DNS (FDNS) and of the LES with that closure (we have also explored matching enstrophy spectra $\hat{Z}(k)=k^2 \hat{E}(k)$; see Section~\ref{sec:Conclusion}. More specifically, we minimize the following loss function with respect to the $C$ value(s):
\begin{eqnarray}\label{eq:loss}
\mathcal{L}=\left\{\text{ln}(\hat{E}^\text{FDNS}) - \text{ln}(\hat{E}^\text{LES}), \Gamma_\text{FDNS}^{-1}\left (\text{ln}(\hat{E}^\text{FDNS}) - \text{ln}(\hat{E}^\text{LES})\right )\right\},
\end{eqnarray}
where  $\{ , \}$ is the Euclidean inner product, $\Gamma_\text{FDNS}$ is the covariance between wavelengths in $\text{ln}(\hat{E}^\text{FDNS})$, and $k=(k_x^2+k_y^2)^{1/2}$ is the wavenumber. 
To obtain the FDNS data and to construct the training and testing datasets, we apply a low-pass sharp-spectral filter with cut-off wavenumber $k_c=L/(2\Delta)=N_\text{LES}/2$ to each DNS snapshot. Therefore, for each case, the FDNS spectrum matches that of the DNS up to $k_c$ (Fig.~\ref{fig:2 DNS and FDNS}). $\hat{E}(k)$ and $\Gamma_\text{FDNS}$ in Eq.~\eqref{eq:loss}, the only information from DNS needed for training, are calculated by averaging the $\hat{E}(k)$ profiles over 100 de-correlated FDNS snapshots to obtain a reliable estimate (See Supplementary Material for more details). The DNS length for this number of FDNS snapshots is within the ``small data'' regime for these cases and not enough for \textit{offline} training of an accurate and stable NN-based closure \cite[]{guan2023learning}. Note that because $\hat{E}(k)$ is rather invariant for flows in statistical equilibrium, we could have used an even shorter DNS from each case, as $\hat{E}(k)$ spectra averaged over just the first 5 DNS snapshots (i.e., a $20\times$ shorter DNS) are virtually indistinguishable from those calculated from all 100 snapshots (see Fig.~\ref{fig:2 DNS and FDNS}), and the optimized parameters are within $1\%$ difference. 

Minimizing loss functions such as Eq.~\eqref{eq:loss} requires performing LES during the optimization process. While traditional data assimilation methods, such as 4D-Variational algorithms~\cite[e.g.,][]{chandramouli20204d}, rely on gradient descent, which requires the explicit calculation of an adjoint model, derivative-free techniques such as EKI~\cite[]{iglesias2013ensemble,chen2012ensemble,bocquet2014iterative} can be used with any solver, which may not be differentiable. In this type of approach, optimizing the model parameters $C$ is framed as maximizing the {\it posterior} probability
\begin{eqnarray}\label{eq:post_prob}
    P(C\mid\text{ln}(\hat{E}^\text{FDNS}),\Gamma_\text{FDNS})=\frac{e^{-\mathcal{L}}}{\zeta(\text{ln}(\hat{E}^\text{FDNS}),\Gamma_\text{FDNS})}P_{prior}(C),
\end{eqnarray}
where $\mathcal{L}$ is the loss as in Eq.~\eqref{eq:loss}, $\zeta(\text{ln}(\hat{E}^\text{FDNS}),\Gamma_\text{FDNS})$ is a normalizing constant, and $P_{prior}(C)$ is the {\it prior} probability density of $C$. Here, as $\mathcal{L}$ is a weighted quadratic, the distribution of $e^{-\mathcal{L}}$ is Gaussian. In the EKI algorithm, the optimization problem is discretized as an interacting particle system. Particles are sampled to describe the {\it prior} distribution of $C$, and then the observed data drives a joint evolution of these particles to ensure that their empirical mean converges quantitatively to the mode of the {\it posterior} probability of $C$ (Eq.~\eqref{eq:post_prob}), solving the inverse problem from the perspective of optimization~\cite[e.g.,][]{iglesias2013ensemble,schneider2020ensemble,cleary2021calibrate}. Similar approaches have been used for LES calibration~\cite{mons2021ensemble,zaki2024turbulence}, though we remark that our EKI algorithm additionally contains adaptive time stepping and regularization to improve convergence and reduce the computational resources. 

In addition to optimization/calibration, we are also interested in quantifying the uncertainty of the {\it posterior} distribution (Eq.~\eqref{eq:post_prob}). In~\cite{mons2021ensemble}, the authors used the empirical covariance of the particles to approximate the covariance of the {\it posterior} distribution of $C$. We did not follow this approach for two reasons: (1) EKI particle spread is often over-confident (``ensemble collapse'') and therefore is not suitable for uncertainty quantification (UQ) without careful inflation procedures (e.g., ensemble Kalman sampler~\cite{garbuno2020interacting} or affine invariant interacting Langevin dynamics~\cite{garbuno2020affine}), and (2) ensemble samplers typically require orders-of-magnitude more LES runs to converge than ensemble optimizers.

Instead, we obtain an approximate UQ by following the approach of CES~\cite{cleary2021calibrate}, which performs sampling (with a Markov-Chain Monte Carlo (MCMC) method) using an emulator trained on the EKI trajectories (all LES results during the EKI training process). The emulator both accelerates and smooths the sampling, and is trained on well-chosen points around the {\it posterior} mode. During the UQ stage, no further LES runs are required.

In addition to TKE spectra, other statistics, such as mean shear stresses, can also be used as targets~\cite[]{mons2021ensemble} in the loss function (Eq.~\eqref{eq:loss}). Here, we find that using TKE spectra as the target provides the best performance and interpretation as it is directly related to the semi-analytical derivation~\cite{guan2025semi}. To quantify the uncertainty of the estimated $C$ value(s), we employ the CES framework \cite[]{cleary2021calibrate}. The framework is well-established and now available as a software~\cite[]{dunbar2024calibrateemulatesample}. 

\section{Results}\label{sec:Results}
\subsection{Online-learned Parameters} 
Table~\ref{tab:1} shows the EKI-optimized constant parameters of Smag, Leith, JHS, and JHL for all 8 cases. The $C_{\text{S}}^{\text{EKI}}$ is around $0.12$ and the same for all cases (within the $\mathcal{O}(0.01)$ uncertainty, see Supplementary Material Table I). This value is smaller than the analytical and empirical value of $0.17$ often used based on previous work with 3D homogeneous isotropic turbulence. As shown below, the EKI-optimized value leads to significant improvements in {\it a-priori} and {\it a-posteriori} metrics over Smag with $C_\text{S}=0.17$ (Smag-0.17). The $C_{\text{L}}^{\text{EKI}}$ is around $0.24$, and again, the same for all cases (within the estimated uncertainty). 
$C_{\text{JHS}}^{\text{EKI}}$ and $C_{\text{JHL}}^{\text{EKI}}$ are also found to be similar across all cases and around $0.215$ and $0.32$, respectively. The backscattering parameter $C_{\text{B}}^{\text{EKI}}$ is the same across all cases and around $0.95$, which is between unity and the previously used value of 0.9. In summary, across these 8 cases, $C_\text{S}$, $C_\text{L}$, $C_\text{JHS}$, $C_\text{JHL}$ and $C_\text{B}$ in these 4 closures are found to be nearly constant. Later in section~\ref{subsec:analytical}, we compare these parameters with those estimated from a semi-analytical derivation, which sheds light on this near universality in the constant value.


For baselines in the {\it a-priori} and {\it a-posteriori tests}, we conduct LES with $\Pi^{\text{SGS}}$ parameterized using Smag-0.17 and using DSmag and DLeith (with positive clipping) implemented following recent studies \cite[]{maulik2016dynamic,guan2023learning}. LES with these 3 ``baseline'' closures and with the 4 EKI-optimized closures are conducted for each case. Each LES run is $100\times$ longer than the duration of the DNS used for the training dataset. As discussed below, for {\it a-priori} tests, we compare the total energy and enstrophy interscale transfers among these closures and FDNS. For {\it a-posteriori} tests, we compare the enstrophy spectra and vorticity PDFs from LES with different closures and from FDNS.  




The global energy and enstrophy transfers are $\langle P_E\rangle=\langle{\Pi}^\text{SGS}\bar{\psi}\rangle$ and $\langle P_Z\rangle=\langle{\Pi}^\text{SGS}\bar{\omega}\rangle$, respectively~\cite[]{thuburn2014cascades, guan2023learning}. Here, $\langle P_E\rangle$ or $\langle P_Z\rangle$ is positive for forward transfer (energy/enstrophy moving to SGS from resolved scales) and negative for backscatter (moving to resolved scales from SGS). $\langle P_E\rangle$ and $\langle P_Z\rangle$ from FDNS and LES for representative cases (1.1, 1.4, 2, 3.3) are shown in Fig.~\ref{fig:table2} (values for all cases are reported in Supplementary Material Tables II and III). It can be observed that for all cases, JHS and JHL have the best agreement with FDNS, and for half of the cases, the values of $\langle P_E\rangle$ and $\langle P_Z\rangle$ closely match those of FDNS. Exceptions are Cases 1.2, 1.3, 1.4, and 3.1, for which JHS and JHL under-predict $\langle P_E\rangle$ by a factor of 2--3; however, they still significantly outperform the optimized Smag and Leith and the baselines, which over-predict $\langle P_E\rangle$  by 1 and even 2 orders of magnitude (thus, these closures are too diffusive). The gain from optimization can be isolated by comparing Smag and Smag-0.17: the former, which uses $C_\text{S} \approx 0.12$, consistently outperforms the baseline in both   $\langle P_E\rangle$ and $\langle P_Z\rangle$ (often by a factor of 2 to 5, or even larger).

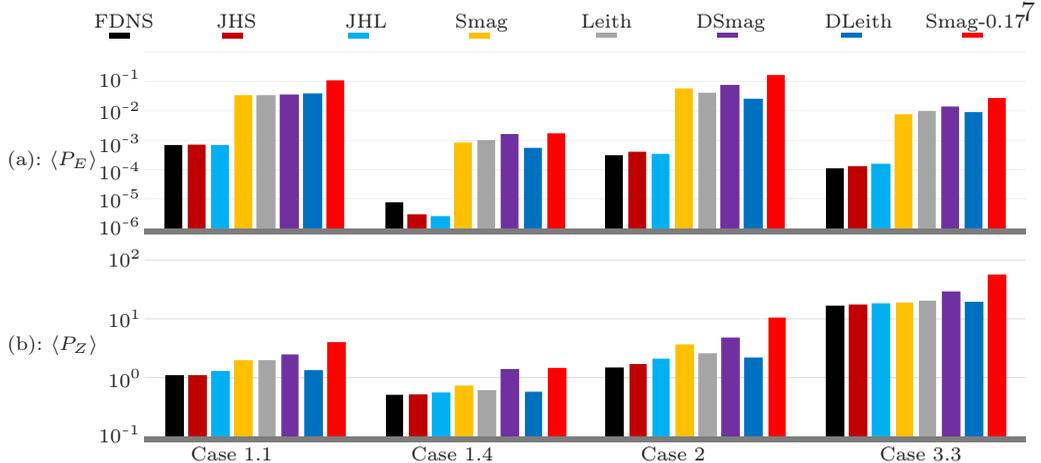
\begin{figure}
\flushright
\vspace{.2in}
 \begin{overpic}[width=0.9\linewidth,height=0.4\linewidth]{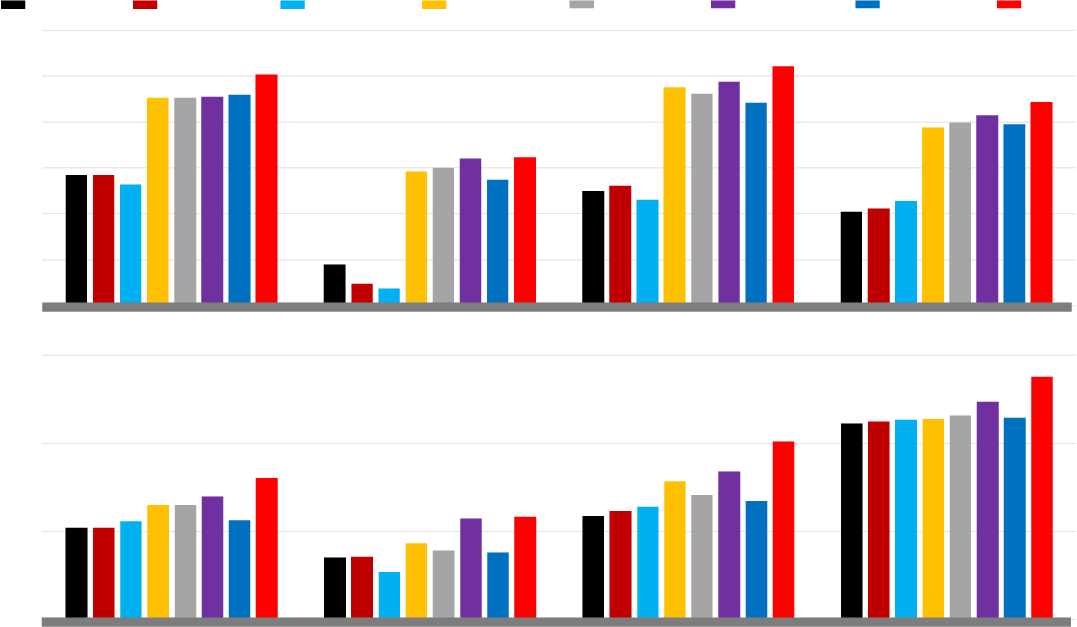}

 \put(-1.5,45){\scriptsize{FDNS}}
 \put(11.8,45){\scriptsize{JHS}}
 \put(25.8,45){\scriptsize{JHL}}
 \put(37.8,45){\scriptsize{Smag}}
 \put(51.5,45){\scriptsize{Leith}}
 \put(64,45){\scriptsize{DSmag}}
 \put(78.0,45){\scriptsize{DLeith}}
 \put(89,45){\scriptsize{Smag-0.17}}
 
 \put(-11,30){\scriptsize (a): $\langle P_E\rangle$}
 \put(-11,10){\scriptsize (b): $\langle P_Z\rangle$}

 \put(9,-2){\scriptsize Case 1.1}
 \put(33,-2){\scriptsize Case 1.4}
 \put(58,-2){\scriptsize Case 2}
 \put(84,-2){\scriptsize Case 3.3}

 \put(-1,22.5){\scriptsize $10^{\shortminus 6}$}
 \put(-1,25.5){\scriptsize $10^{\shortminus 5}$}
 \put(-1,28.5){\scriptsize $10^{\shortminus 4}$}
 \put(-1,31.5){\scriptsize $10^{\shortminus 3}$}
 \put(-1,35){\scriptsize $10^{\shortminus 2}$}
 \put(-1,38.5){\scriptsize $10^{\shortminus 1}$}

 \put(-1,0){\scriptsize $10^{\shortminus 1}$}
 \put(0,6){\scriptsize $10^{0}$}
 \put(0,12.5){\scriptsize $10^{1}$}
 \put(0,19){\scriptsize $10^{2}$}



 \end{overpic}
  \vspace*{2mm}
 \caption{\footnotesize Interscale transfers in representative cases calculated {\it a-priori} for the same FDNS samples (note the logarithmic scale). EKI-optimized closures (JHS, JHL, Smag, and Leith) are compared against baselines (DSmag, DLeith, and Smag-0.17) and the truth (FDNS). Values for all 8 cases are shown in Supplementary Material Tables II and III.}
 \label{fig:table2}
 \vspace*{2mm}
\end{figure}
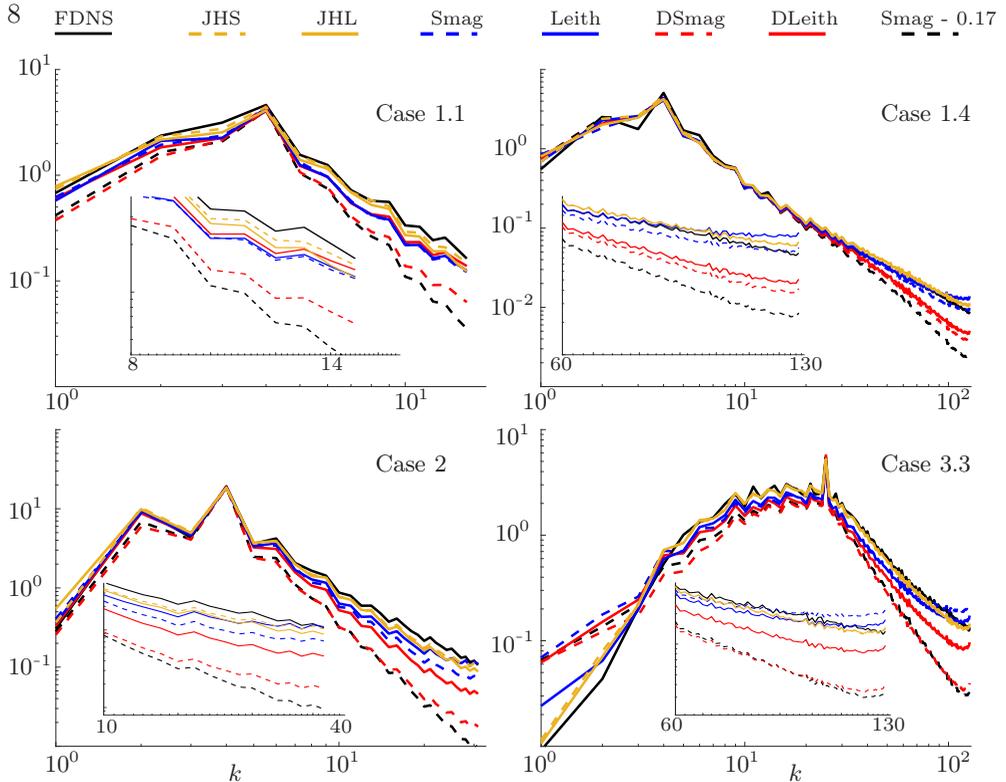
\begin{figure}
\vspace{0.2in}
 \centering
 \begin{overpic}[width=0.9\linewidth,height=0.7\linewidth]{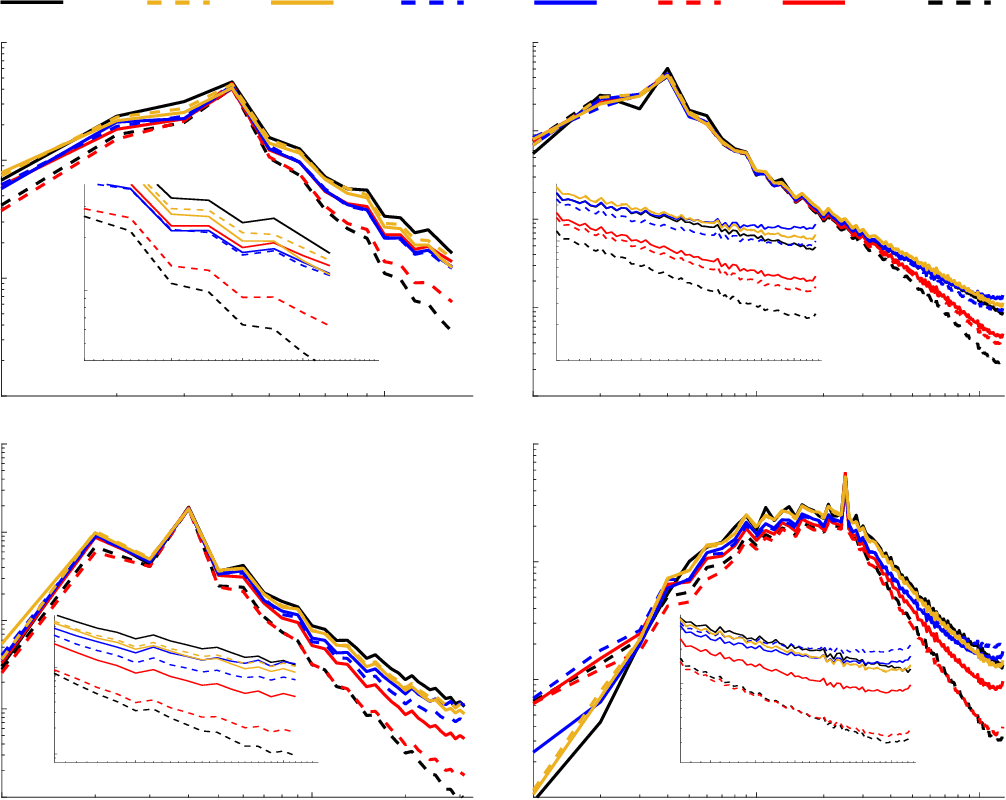}

 \put(0,78.5){\scriptsize{FDNS}}
 \put(16,78.5){\scriptsize{JHS}}
 \put(28.5,78.5){\scriptsize{JHL}}
 \put(41,78.5){\scriptsize{Smag}}
 \put(54,78.5){\scriptsize{Leith}}
 \put(65.5,78.5){\scriptsize{DSmag}}
 \put(78,78.5){\scriptsize{DLeith}}
 \put(90,78.5){\scriptsize{Smag - $0.17$}}

 \put(35,68){{Case 1.1}}
 \put(90,68){{Case 1.4}}
 
 \put(35,30){{Case 2}}
 \put(90,30){{Case 3.3}}

 \put(-1,36.8){\footnotesize{$10^0$}}
 \put(37,36.8){\footnotesize{$10^1$}}

 \put(51,36.8){\footnotesize{$10^0$}}
 \put(73,36.8){\footnotesize{$10^1$}}
 \put(96,36.8){\footnotesize{$10^2$}}

 \put(19,-3.5){\footnotesize{$k$}}
 \put(-1,-2.5){\footnotesize{$10^0$}}
 \put(29.5,-2.5){\footnotesize{$10^1$}}
 
 \put(80,-3.5){\footnotesize{$k$}}
 \put(51,-2.5){\footnotesize{$10^0$}}
 \put(73,-2.5){\footnotesize{$10^1$}}
 \put(96,-2.5){\footnotesize{$10^2$}}

 \put(-5,50){\footnotesize{$10^{\shortminus1}$}}
 \put(-4,61.5){\footnotesize{$10^{0}$}}
 \put(-4,73){\footnotesize{$10^{1}$}}

 \put(-5,8){\footnotesize{$10^{\shortminus1}$}}
 \put(-5,16.5){\footnotesize{$10^{0}$}}
 \put(-5,25){\footnotesize{$10^{1}$}}

 \put(47.5,47){\footnotesize{$10^{\shortminus2}$}}
 \put(47.5,56){\footnotesize{$10^{\shortminus1}$}}
 \put(47.8,64.3){\footnotesize{$10^{0}$}}

 \put(47.5,10.5){\footnotesize{$10^{\shortminus1}$}}
 \put(48,22.5){\footnotesize{$10^{0}$}}
 \put(48,33){\footnotesize{$10^{1}$}}

 \put(8,41){\scriptsize{$8$}}
 \put(29,41){\scriptsize{$14$}}
 
 \put(4.5,1.7){\scriptsize{$10$}}
 \put(30,1.7){\scriptsize{$40$}}
 
 \put(54,41){\scriptsize{$60$}}
 \put(80,41){\scriptsize{$130$}}
 
 \put(66,2){\scriptsize{$60$}}
 \put(89,2){\scriptsize{$130$}}

 \end{overpic}
  \vspace*{2mm}
 \caption{\footnotesize Enstrophy spectra ($\hat{Z}$) from FDNS and LES for representative cases (see Supplementary Material for other cases). The insets magnify the high-$k$ part of spectra.}
 \label{fig:3}
 \vspace*{2mm}
\end{figure}

Next, we examine the enstrophy spectra, $\hat{Z}(k)$, and the PDF of vorticity, $\mathcal{P}({\omega})$. Figure~\ref{fig:3} shows $\hat{Z}(k)$ for representative cases. Smag-$0.17$ and DSmag, consistent with $\langle P_Z\rangle$ in Fig.~\ref{fig:table2}, have excessive enstrophy dissipation and under-predict, by about an order of magnitude, $\hat{Z}$ compared to FDNS at the smallest scales. Among the baselines, DLeith performs the best in matching the FDNS, but it is still not comparable to EKI-optimized models. All the EKI-optimized closures work well in matching the FDNS $\hat{Z}$, which is not surprising given that their target was TKE spectra. The one exception is Case 3.3, where only optimized JH closures (and to a lesser degree Leith) can match the FDNS $\hat{Z}$ in the largest scales (small $k$).

\begin{figure}
\vspace{.2in}
 \centering
 \begin{overpic}[width=0.9\linewidth,height=0.7\linewidth]{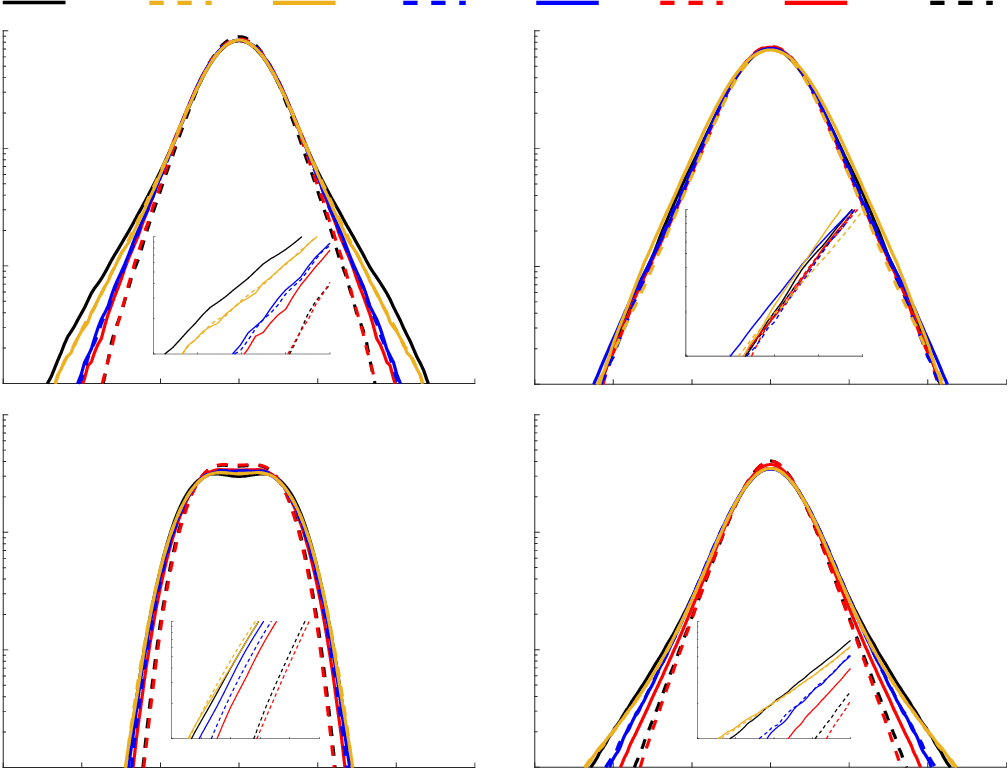}

 \put(0,79){\scriptsize{FDNS}}
 \put(16,79){\scriptsize{JHS}}
 \put(28.5,79){\scriptsize{JHL}}
 \put(41,79){\scriptsize{Smag}}
 \put(54,79){\scriptsize{Leith}}
 \put(66,79){\scriptsize{DSmag}}
 \put(78,79){\scriptsize{DLeith}}
 \put(90,79){\scriptsize{Smag - $0.17$}}

 \put(35,70){{Case 1.1}}
 \put(90,70){{Case 1.4}}
 
 \put(35,30){{Case 2}}
 \put(90,30){{Case 3.3}}

 \put(16,-4){\small{$\bar{\omega}/\sigma_{\omega}$}}
 \put(-1,-2){\footnotesize{$\shortminus6$}}
 \put(23.2,-2){\footnotesize{$0$}}
 \put(46,-2){\footnotesize{$6$}}

 \put(70,-4){\small{$\bar{\omega}/\sigma_{\omega}$}}
 \put(52,-2){\footnotesize{$\shortminus6$}}
 \put(75.8,-2){\footnotesize{$0$}}
 \put(99,-2){\footnotesize{$6$}}

 \put(-5.0,-1){\footnotesize{$10^{\shortminus4}$}}
 \put(-5.0,11.5){\footnotesize{$10^{\shortminus3}$}}
 \put(-5.0,23.5){\footnotesize{$10^{\shortminus2}$}}
 \put(-5.0,35.0){\footnotesize{$10^{\shortminus1}$}}

 \put(-5.0,38.5){\footnotesize{$10^{\shortminus4}$}}
 \put(-5.0,50.5){\footnotesize{$10^{\shortminus3}$}}
 \put(-5.0,62.0){\footnotesize{$10^{\shortminus2}$}}
 \put(-5.0,73.5){\footnotesize{$10^{\shortminus1}$}}

 \put(14,40.2){\scriptsize{$\shortminus5$}}
 \put(22.5,40.2){\scriptsize{$\shortminus4$}}
 \put(31,40.2){\scriptsize{$\shortminus3$}}

 \put(67,40.2){\scriptsize{$\shortminus5$}}
 \put(75,40.2){\scriptsize{$\shortminus4$}}
 \put(83.5,40.2){\scriptsize{$\shortminus3$}}
 
 \put(15.5,1.4){\scriptsize{$\shortminus3$}}
 \put(30,1.4){\scriptsize{$\shortminus2$}}

 \put(67.5,1.4){\scriptsize{$\shortminus5$}}
 \put(75,1.4){\scriptsize{$\shortminus4$}}
 \put(82.5,1.4){\scriptsize{$\shortminus3$}}
 
 \end{overpic}
  \vspace*{5mm}
 \caption{\footnotesize PDFs of $\bar{\omega}$ normalized by the standard deviation of FDNS ($\sigma_\omega$) for representative cases (see Supplementary Material for other cases). The insets zoom into the left tails. Kernel estimator with a bandwidth of 1 is used to generate the PDFs.}
 \label{fig:4}
 \vspace*{2mm}
\end{figure}

Figure~\ref{fig:4} shows PDF $\mathcal{P}(\bar{\omega})$ for representative cases. Except for Cases 1.3 and 1.4 for which LES with all closures match the FDNS' PDF well (down to the tails), for all other cases, LES with optimized JH closures clearly outperform the baselines and optimized Smag and Leith in matching the FDNS $\mathcal{P}$, especially at the tails (rare, extreme events). In fact, except for Case 1.1, which involves the largest $N_\text{DNS}/N_\text{LES}=32$, LES with optimized JH matches the FDNS' $\mathcal{P}$ well. LES with optimized Smag and Leith also outperform the baselines (see, e.g., 3.3), although for Cases 1.1, 1.2, and 3.1, DLeith shows comparable performance. 

\subsection{Comparison with Semi-Analytical Estimates} \label{subsec:analytical}
To fully understand the EKI-optimized parameters, we compare the online-learned estimated with predictions from a semi-analytical derivation that was recently introduced in Guan and Hassanzadeh~\cite{guan2025semi}. The semi-analytical derivation uses the  interscale enstrophy transfer $\eta=\langle\bar{\omega}\Pi\rangle$~\cite[]{thuburn2014cascades,guan2023learning} and the TKE scaling law $\hat{E}(k)=A\eta^{2/3}k^{-3}$~\cite[]{kraichnan1967inertial,leith1968diffusion,batchelor1969computation} to obtain:
\begin{eqnarray}\label{eq:CS}
\color{black} C_\text{S}=(A^3/2)^{-1/4} \pi^{-1} (ln(k_c))^{-1/4},
\end{eqnarray}
\begin{eqnarray}\label{eq:CL}
C_\text{L}=1/(\pi A^{1/2}),
\end{eqnarray}
and
\begin{eqnarray}\label{eq:CHL_CB}
    C_\text{JHL}&=&(A/2)^{-1/4}\pi^{-1} (1-C_\text{B}/ln(k_c))^{-1/6}.
\end{eqnarray}
$A$ in these equations can be obtained by linearly fitting the scaling law in the inertial region ($[k_f+1,k_c]$) of the TKE spectra from a few DNS snapshots (see Table.~\ref{tab:1} for the estimated values). In the fitting process, $\eta$ is obtained by averaging the value for the FDNS snapshots. The near universality of the online-learned $C_\text{L}$, $C_\text{S}$, and $C_\text{JHL}$ can be explained by the consistency in $A\approx 1.8-1.9$ across most cases (except for Case 2) and the weak dependence on $k_c$, see Eqs.~\eqref{eq:CS} and~\eqref{eq:CHL_CB}. The value of $A$ for 2D homogeneous isotropic turbulence here is close to the one predicted using the renormalization-group analysis, i.e., $A=1.923$~\cite[e.g.,][]{olla1991renormalization,nandy1995mode}.
In fact, for most 2D homogeneous isotropic turbulence, $A$ is found empirically to be around $1.0-2.0$~\cite[][]{smith1993bose,gotoh1998energy,lindborg2010testing,boffetta2012two,gupta2019energy}. The difference in $A$ for Case 2 is likely due to the $\beta$ effect and the resulting strong anisotropy. Nevertheless, once $A$ is determined, e.g., empirically using only the TKE spectra of a few DNS snapshots, the semi-analytical derivation can be used to estimate the model parameters. Furthermore, the consistency between the EKI-optimized parameters and the semi-analytical prediction within $\mathcal{O}(0.01)$ uncertainty suggests that these parameters are learned from data such that they match the enstrophy interscale transfer ($\eta$) of LES to FDNS.

\section{Summary and Conclusion}\label{sec:Conclusion}
We use the CES framework with EKI to learn \textit{online} the parameters of four well-known physics-based models for 8 setups of 2D geophysical turbulence, and interpret them with a new semi-analytical derivation. The objective of the learning process is to match the TKE spectrum $\hat{E}$ of LES with the optimized closure with the spectrum of FDNS. The latter can be obtained from a short DNS run, making the approach \textit{data-efficient}. Note that we have also explored using the enstrophy spectrum $\hat{Z}$ of FDNS as the target for learning $C_\text{S}$ and $C_\text{L}$ for Case 3.1, and found similar constant parameters (1.2 vs. 1.3 and 2.4 vs. 2.4), which is not surprising given the simple relationship between $\hat{Z}$ and $\hat{E}$. The estimated parameters of each closure are found to be nearly constant across the 8 cases, suggesting that the optimized closures are \textit{generalizable}. We further analyze these parameters using turbulence theory and scaling analysis and verify that the parameters are optimized to match the FDNS interscale enstrophy transfer. Such analysis provides further insight into these already \textit{interpretable} closures. Furthermore, this comparison also provides further support for the the semi-analytical derivation, which involves several assumptions and simplifications. The semi-analytical derivation has its own significance: to calculate the model parameters, e.g., $C_\text{S}, C_\text{L}, \text{and } C_\text{JHL}$, it only requires estimating $A$ from a short high-resolution simulation, without any deep learning or online learning techniques.

A number of metrics are used to quantify the performance of the optimized closures. The {\it a-priori} metrics involving mean interscale energy and enstrophy transfers (between the SGS and resolved flow) show major improvements resulting from optimization. In particular, the backscattering model (JH) demonstrates significantly better ability in capturing these transfers compared to baselines and optimized eddy-viscosity models. {\it A-posteriori} tests, focused on examining the enstrophy spectra at the largest and smallest length scales and the tails of the vorticity PDFs, again show major improvements gained from the optimization. Overall, the optimized eddy-viscosity closures are found to work well (and better than the baselines) except for cases in which the LES resolution is low compared to the forcing length scale (i.e., small $k_c/k_f$). For such cases, the eddy-viscosity closures have structural errors, and optimization alone (which only addresses parameteric errors) cannot further improve the LES. The JH closure, on the other hand, leads to LES that matches FDNS closely based on all metrics and for all cases, further showing the importance of combining theoretical advances and data-driven techniques to develop data-efficient frameworks for accurate, interpretable, and generalizable closures. The next steps in this work should include exploring cases with more extreme regimes (e.g., higher $Re$) and different dynamics, and investigating more complex test cases (e.g., ocean general circulation models).

\section*{Acknowledgments}
We thank Karan Jakhar, Malte Jansen, and Rambod Mojgani for insightful discussions. This work was supported by ONR award N000142012722, NSF grants OAC-2005123 and AGS-1835860, and by Schmidt Sciences, LLC. Computational resources were provided by ACCESS (ATM170020) and CISL (URIC0004). Codes/data are available at~\url{https://github.com/envfluids/2D-DDP} and~\url{https://clima.github.io/CalibrateEmulateSample.jl/}.

\bibliography{EKI_PRR}

\end{document}